\documentclass[
               ]{jetp} 

\twocolumn 
\usepackage{stmaryrd}
\usepackage{capt-of}
\usepackage{xcolor}

\begin{document}
\English

\title{Universal phase diagram and scaling functions of imbalanced Fermi gases} 



\affiliation{Physik-Department, Technische Universit\"at M\"unchen, D-85748 Garching, Germany}
\author{Bernhard}{Frank} 
\author{Johannes}{Lang}
\author{Wilhelm}{Zwerger} 
\email{zwerger@ph.tum.de}

\abstract{We discuss the phase diagram and the universal scaling functions 
of attractive Fermi gases at finite imbalance. The existence of a quantum 
multicritical point for the unitary gas at vanishing chemical potential $\mu$ and
effective magnetic field $h$, first discussed by Nikoli\'{c} and Sachdev, 
gives rise to three different phase diagrams, depending on whether
the inverse scattering length $1/a$ is negative, positive or zero. Within
a Luttinger-Ward formalism, the phase diagram and pressure
of the unitary gas is calculated as a function of the dimensionless scaling
variables $T/\mu$ and $h/\mu$. The results indicate that beyond 
the Clogston-Chandrasekhar limit at $(h/\mu)_c\simeq 1.09$, the unitary gas 
exhibits an inhomogeneous superfluid phase with FFLO order that can reach critical temperatures near unitarity of $\simeq 0.03\, T_F$.}

\maketitle


\section{Introduction}




The experimental realization of stable,  two-component 
Fermi gases near a Feshbach resonance, where the 
magnitude of the scattering length $a$ can be tuned far beyond 
the average interparticle spacing, allows to explore the
crossover from a Fermi superfluid of the BCS-type to a Bose-Einstein condensate 
of tightly bound pairs 
(for recent reviews of this subject see~\cite{rand14},~\cite{zwie14} and~\cite{zwer14varenna}). 
Of particular interest in this context is the unitary regime $k_F|a|\gg 1$. 
At first sight, the point at which the scattering length diverges is not expected to show any
special features, because the ground state is a superfluid on both sides of the unitary point. 
This argument, however, misses a number of essential features: 
As pointed out by Nishida and Son~\cite{nish07CFT}, the unitary gas realizes a non-relativistic field theory which is both 
scale and conformally invariant. The additional symmetries have a number of nontrivial 
consequences, like  a vanishing bulk viscosity~\cite{son07bulk} or a breathing mode at twice 
the trap frequency in the presence of a harmonic confinement~\cite{wern06unitary}. The latter is due to a 
hidden SO$(2,1)$ symmetry first discussed for Bose gases in two dimensions~\cite{pita97symmetry}.
Moreover, as shown by Nikoli\'{c} and Sachdev~\cite{niko07renorm}, 
the unitary gas at zero density realizes a quantum multicritical point. It 
separates the onset transition from the vacuum to a finite density superfluid into 
two regimes where the flow is towards a weakly interacting gas of either Fermions or Bosons.
The thermodynamics of a Fermi gas near unitarity is therefore
governed by a novel strong coupling fixed point and associated universal scaling functions. \\

In the following, we will discuss the consequences of 
this basic insight for the phase diagram and scaling functions of 
imbalanced Fermi gases at finite values of
the chemical potential difference  $h=(\mu_{\shortuparrow}-\mu_{\shortdownarrow})/2$
(for reviews of this subject see~\cite{chev10review} or the chapters by
Chevy and Salomon or by Recati and Stringari in~\cite{zwer12book}). 
It turns out that - in combination with input from existing and new microscopic 
results - this approach provides a quantitative description of their thermodynamic properties.
In particular, we will determine the universal phase
diagram at zero temperature as a function of the effective magnetic field $h$ and the 
chemical potential $\mu$. At unitarity, scale invariance implies that the phase boundaries are simply straight
lines with slopes that are universal numbers. Away from unitarity, the phase diagram at arbitrary values of 
the scattering length collapses into just two different universal diagrams by scaling the chemical 
potential $\bar{\mu}=\mu/(\varepsilon_b/2)$ with the characteristic two-particle energy $\varepsilon_b=\hbar^2/(ma^2)$.
The diagram covering negative values of the scattering length has the form known from BCS-theory
in the presence of a Zeeman field which couples to the spin imbalance. In particular, it exhibits an 
FFLO type superfluid 
\footnote{As emphasized by Radzihovsky~\cite{radz11FFLO}, FFLO is a pair-density wave rather than a genuine supersolid because there is only a single 
order parameter, not two independent ones.}
with a spatially varying order parameter proportional to $\cos{(\mathbf{Q}\cdot\mathbf{x})}$ in the regime $1/\sqrt{2}<h/\Delta<0.754$, 
where $\Delta$ is the quasiparticle gap at $h=0$~\cite{FF64,LO64}. An FFLO phase is also
present at unitarity and it extends to positive values of the scattering length. For $a>0$, 
there is again a universal phase  diagram $\bar{h}=h/(\varepsilon_b/2)$ versus $\bar{\mu}$,  which is fundamentally different from the one at $a<0$. 
The fact that the ground state of the unitary gas exhibits FFLO order beyond the Clogston-Chandrasekhar limit at $(h/\mu)_c\simeq 1.09$,
where a finite imbalance $s=n_{\shortuparrow}-n_{\shortdownarrow}\ne 0$ sets in, is consistent 
with a field-theoretical analysis by Son and Stephanov~\cite{son06imbalance} and also with numerical studies e.g.~\cite{bulg08lo}. 
It is supported below by a calculation of the phase diagram and scaling functions within a Luttinger-Ward
approach, extending earlier results~\cite{haus07bcsbec} to finite values of the imbalance. The range of temperatures, where 
FFLO order appears is restricted to temperatures smaller than $\simeq 0.03\,T_F$,
making its observation rather challenging, even with the availability of confining potentials 
for ultracold Fermi gases which are flat over considerable length scales~\cite{mukh17}. 

\section{FESHBACH RESONANCES AND LOW DENSITY LIMIT}

The regime of strong interactions in dilute, ultracold gases 
can be reached via Feshbach resonances~\cite{chin10feshbach}.
In the case of magnetically tunable resonances, a standard 
parametrization of the scattering length 
near a particular resonance at $B=B_0$ is given by 
\begin{equation}
\label{eq:FBs-length}
a(B)=a_{\rm bg}\left(1-\frac{\Delta B}{B-B_0}\right)\to -\frac{\hbar^2}{mr^{\star}\Delta\mu (B-B_0)}+\ldots\, .
\end{equation}
Here, $a_{\rm bg}$ is the off-resonant background scattering length
in the absence of the coupling to the closed channel, while 
$\Delta B$ describes the width of the resonance expressed in magnetic field units. 
The resonant contribution to the scattering length is inversely proportional to the 
detuning $\nu(B)=\Delta\mu (B-B_0)$, where $\Delta\mu$ is the difference in magnetic
moments between the open and closed channel, which can be either positive or negative. 
The relation $1/a=-mr^\star\nu(B)/\hbar^2+\ldots$ near the resonance  defines a positive, 
intrinsic length scale $r^{\star}>0$. Characteristic values of the background scattering length
$a_{\rm bg}$ are of the order of the van der Waals length $ l_{\rm vdW}=\left(mC_6/\hbar^2\right)^{1/4}/2$,
which is determined by the strength $C_6$ of the attractive interaction at large distances~\cite{grib93scatt}.
In the regime of interest here, this is much smaller than the resonant contribution.\\

A microscopic description of interactions near a Feshbach resonance,
which neglects the background contribution to the scattering length, 
can be obtained within a two-channel model.
For a two-component Fermi gas, the effective Hamiltonian is 
\begin{wide}
\begin{align}
\begin{split}
\label{eq:BFM}
\hat{H}_{\rm Feshbach}=\int_{\mathbf{x}}\,\Bigg[& \sum_{\sigma}\hat{\psi}_{\sigma}^{\dagger}(\mathbf{x})
\big(-\frac{\hbar^2}{2m}\nabla^2\big)\hat{\psi}_{\sigma}(\mathbf{x})\; +\hat{\Phi}^{\dagger}(\mathbf{x})\big(-\frac{\hbar^2}{4m}\nabla^2+\nu_{c}(B)\big)
\hat{\Phi}(\mathbf{x}) \\
&+ \tilde{g}\,\int_{\mathbf{x}'}\,  \chi(|\mathbf{x}-\mathbf{x'}|) \Big( \hat{\Phi}^{\dagger}(\frac{\mathbf{x}+\mathbf{x'}}{2})\,
 \hat{\psi}_{\shortuparrow}(\mathbf{x}) \hat{\psi}_{\shortdownarrow}(\mathbf{x'})+\rm{h.c.}\Big)\Bigg]\,.
\end{split}
\end{align}
\end{wide}

Here, the fermionic field operators $\hat{\psi}_{\sigma}(\mathbf{x})$ 
describe atoms in the open channel, with a formal 
spin variable $\sigma=\shortuparrow,\shortdownarrow$ distinguishing 
two different hyperfine states. The bound
state in the closed channel is denoted by the bosonic operator $\hat{\Phi}$.
Its energy $\nu_{c}(B)$ measures the detuning of the bare closed channel 
bound state with respect to two atoms at zero energy. 
The coupling is characterized by a strength $\tilde{g}$ and a form factor 
$\chi(\mathbf{x})$, which only depends on the magnitude $|\mathbf{x}-\mathbf{x'}|$ of the 
distance between two atoms in the open channel and is normalized by $\int_{\mathbf{x}}\chi(\mathbf{x})=1$.
In the following,  the two-channel model of Eq.~(\ref{eq:BFM}) will be replaced by an effective direct 
interaction between the Fermions, 
which is obtained by integrating out  the bosonic field $\hat\Phi$. 
The effective scattering amplitude of two atoms with momenta $\pm\mathbf{k}$ in their center of mass frame is given by
\begin{equation}
\label{eq:fkexact}
f(k)=\frac{m}{4\pi\hbar^2}\,\tilde{g}^2\chi^2(k)\, \mathcal{G}_\Phi(E=\frac{\hbar^2k^2}{m},\mathbf{Q}=0)\, .
\end{equation}
Its dependence on the momentum $k$ is determined by the inverse propagator 
\begin{equation}
\label{ef3_FullG}
\mathcal{G}_\Phi^{-1}(E,\mathbf{Q}=0) = -E+\nu_c(B)+\frac{m\tilde{g}^2}{\hbar^2}\,\int_q\frac{\chi^2(q)}{k^2-q^2 +i0}
\end{equation}
of the field $\hat\Phi$ at vanishing total momentum $\mathbf{Q}=0$ and energy $E=\hbar^2k^2/m$.
Expanding $f(k)$ at low energies, the scattering amplitude has the standard form
\begin{equation}
\label{eq:s-ampl}
f(k)=\frac{1}{k\cot{\delta_0(k)}-ik}\,\to\,
\frac{1}{-1/a+r_ek^2/2+\dots -ik}\, .
\end{equation}
The associated scattering length $a$ and effective range $r_e$ are given by~\cite{schm12efimov}
\begin{equation}
       \frac{1}{a}\!=\!-\frac{mr^\star}{\hbar^2}\nu_{c}(B)+\frac{1}{2\sigma}
    \;\; {\rm and} \;\;
    r_e \!=\!-2r^\star\!+3\sigma\!\left(\!1-\frac{4\sigma}{3a} \!\right).
      \label{eq:FB-range}
\end{equation}
The length $r^\star=4\pi\hbar^4/(m^2\tilde{g}^2)$, which is inversely proportional to the square $\tilde{g}^2$ of the 
Fesh\-bach coupling, turns out to coincide with the parameter introduced above.  
The second length scale $\sigma$, which arises from a Lorentzian cutoff $\chi(\mathbf{k})=1/\left(1+(k\sigma)^2\right)$ 
in momentum space, is essentially equal to the van der Waals length $ l_{\rm vdW}$.
This follows by expanding the detuning $\nu_{c}(B)=\Delta\mu(B-B_{c})$
of the closed channel molecular state to linear order around a bare resonance position $B_{c}$.
The inverse scattering length $1/a(B)=-mr^\star\nu(B)/\hbar^2$ then indeed has the form given
in~(\ref{eq:FBs-length}). The resonance position is, however, shifted from its bare value by 
\begin{equation}
\Delta\mu\left( B_0-B_{c}\right)=\frac{\hbar^2}{2mr^\star\sigma}\, .
\label{eq:shift}
\end{equation} 
 This shift has been calculated within a microscopic description of the
Feshbach coupling using multichannel quantum defect theory~\cite{gora04shift}. 
Comparison with this result yields the identification of the effective range $\sigma$ of 
$\chi(\mathbf{x})$ with the mean scattering length $\bar{a}=0.956\,l_{\rm vdW}$ \cite{schm12efimov}.\\

Depending on the value of the resonance strength parameter  
$s_{\rm res}=\bar{a}/r^\star$,
Feshbach resonances can be classified as either open or closed channel dominated~\cite{chin10feshbach}.
In the open channel dominated case $s_{\rm res}\gg 1$,
the effective range $r_e\to 3\,\sigma=3\, \bar a$ essentially coincides
with the value $r_e\to 2.92\, \bar a$ obtained for a single-channel potential with a $1/r^6$ tail
in the relevant regime where $|a|\gg\bar{a}$~\cite{flam99scatt}.
In the case of closed channel dominated resonances with $s_{\rm res}\ll 1$, in turn,  Eq.~(\ref{eq:FB-range})
gives rise to a negative effective range $r_e\to -2r^\star$, whose magnitude is large compared to the 
characteristic scale set by $\bar{a}\simeq l_{\rm vdW}$
\footnote{Note that in many publications, the form factor $\chi(\mathbf{x})$ in~(\ref{eq:BFM}) is replaced by a delta function,
which leads to the incorrect result $r_e=-2r^\star$ for \emph{all} Feshbach resonances.}.
Now, in order to obtain a thermodynamic potential for a Fermi gas
at finite density, where the interaction is completely specified by the 
scattering length alone, it is necessary that the effective two-body scattering amplitude~(\ref{eq:fkexact})  
is of the idealized form $f(k)=-a/(1+ika)$ of a contact interaction 
at all relevant wave vectors. This requires the effective range $r_e$ to be negligible at $k\simeq k_F$,
which is valid in the limit $k_F|r_e|\to 0$. It is obvious that this limit is eventually always reached at low 
enough densities. For closed channel dominated resonances, the condition $k_F|r_e|\ll 1$ 
coincides with the requirement that the fraction
 \begin{equation}
Z=\frac{1}{N/2}\int _{\mathbf{R}}\;\langle\hat{\Phi}^{\dagger}(\mathbf{R})\hat{\Phi}(\mathbf{R})\rangle\,\simeq k_Fr^{\star}/2\ll 1
\label{eq:Nb}
\end{equation} 
of closed channel molecules near the resonance is small compared to one~\cite{wern09closed,zwer14varenna}, a condition which is often called 
the 'broad resonance' limit
\footnote{This must be distinguished carefully from the notion of open or closed channel dominated resonances discussed above,
which is defined at the level of two-body interactions, independent of the Fermion density.}.
For the relevant case of open channel dominated resonances, where $r^\star\ll\bar{a}$,
the condition $k_F|r_e|\ll 1$ requires, however that $k_F\bar{a}\ll 1$, which is much more restrictive but still well obeyed
in practice. A specific example is the Feshbach resonance in $^6$Li near $832\,$G, where $s_{\rm res}=\bar{a}/r^{\star}\simeq 59$
and $\bar{a}\simeq 30\, a_B$~\cite{chin10feshbach}. For typical values of the Fermion density, the closed channel fraction 
$Z\simeq k_F\bar{a}/(2s_{\rm res})\simeq10^{-4}$ near $B_0$ is then negligible
since both $k_F\bar{a}$ and $1/(2s_{\rm res})$ are very small.\\

In practice, it is only for open channel dominated Feshbach resonances, where 
the multi\-critical fixed point of a unitary gas with zero range interactions, which will be discussed below,
is accessible.  Indeed, in the closed channel case, the broad resonance condition~(\ref{eq:Nb}) implies
that the cutoff  $\varepsilon^{\star}=\hbar^2/m(r^{\star})^2$ in energy of the effective Fermi-Fermi interaction 
due to finite range effects with $r_e=-2r^\star$ is much larger than the Fermi energy $\varepsilon_F$. This is necessary 
for universality, which requires that $\varepsilon_F$ and $\varepsilon_b=\hbar^2/ma^2$ are the only energy scales in the problem.  
The condition that the resonant contribution to the scattering length is much larger than typical 
background values $\bar{a}$, however, requires $|\nu(B)|/\bar{E}\ll s_{\rm res}$, where $\bar{E}=\hbar^2/m\bar{a}^2$
is the van der Waals energy, of order $670\,$MHz for $^6$Li. For closed channel dominated Feshbach resonances,
where $s_{\rm res}\ll 1$, the range of detunings $\nu(B)$ where $|a|\gg\bar{a}$ is thus very narrow. 
In the open channel dominated case, in turn, the condition $|\nu(B)|/\bar{E}\ll s_{\rm res}$
is obeyed over a wide range of detunings because $s_{\rm res}\gg 1$. In this context, it
is also important to emphasize that the case $\gamma\simeq 1/(k_Fr^\star)\ll 1$ of a 'narrow' Feshbach resonance, as 
used e.g. in~\cite{shee06mag}, leads to a completely different universality class. For such resonances, 
the bosonic field $\hat{\Phi}$ in the basic Hamiltonian~(\ref{eq:BFM}) may be replaced by a c-number gap function $\Delta(\mathbf{x})$ via
\begin{equation}
\tilde{g}\,\int_{\mathbf{x}'}\,  \chi(|\mathbf{x}-\mathbf{x'}|)\, \hat{\Phi}^{\dagger}(\frac{\mathbf{x}+\mathbf{x'}}{2})\;\to\;\Delta(\mathbf{x})\, .
\label{eq:replace}
\end{equation}
The  two-channel model is thus reduced to the exactly solvable BCS Hamiltonian.  
The replacement~(\ref{eq:replace})
is legitimate, because for $s_{\rm res}\ll 1$ the closed channel state
is responsible for the interaction between the Fermions in the open channel
but is unaffected by their condensation, similar to phonons in
conventional superconductors. The associated description corresponds to
taking a \emph{high density} limit. It gives rise to the well-known BCS-universality,
which is quite different from the universality discussed below~\cite{zwer14varenna}.\\

In the low density limit $k_F|r_e|\to 0$, the basic two-channel model~(\ref{eq:BFM}) can be replaced 
by an effective Hamiltonian
\begin{wide}
\begin{equation}
\label{eq:H-direct}
\hat{H}=\int_{\mathbf{x}}\,\Big[ \sum_{\sigma}\hat{\psi}_{\sigma}^{\dagger}(\mathbf{x}) \big(-\frac{\hbar^2}{2m}\nabla^2\big)\hat{\psi}_{\sigma}(\mathbf{x})\; +
\bar{g}(\Lambda)\,\hat{\psi}_{\shortuparrow}^{\dagger}(\mathbf{x}) \hat{\psi}_{\shortdownarrow}^{\dagger}(\mathbf{x}) \hat{\psi}_{\shortdownarrow}(\mathbf{x}) \hat{\psi}_{\shortuparrow}(\mathbf{x})\Big]\, ,
\end{equation}
\end{wide}
which involves a direct delta function interaction $V(\mathbf{x})\to\bar{g}(\Lambda)\,\delta(\mathbf{x})$ between Fermions.
This Hamiltonian needs a cutoff in energy due to the finite range of the physical interactions. For open channel dominated
Feshbach resonances, this cutoff is the van der Waals energy $\bar{E}$. Formally, the delta function 
interaction leads to a vanishing scattering amplitude if $\bar{g}$ is kept finite.    
The coupling constant $\bar{g}(\Lambda)$ must therefore be adjusted properly to 
give rise to a non-vanishing scattering length $a$. Taking the limit $k\to 0$ in the two-body Lippmann-Schwinger 
equation and defining the physical coupling constant $g=4\pi\hbar^2 a(B)/m$, this implies the relation 
\begin{align}
\frac{1}{\bar{g}}=\frac{1}{g}-\int_{q<\Lambda}\frac{1}{2\varepsilon_q}\, .
\label{eq:gbar}
\end{align}
Here $\varepsilon_q=\hbar^2q^2/2m$ is the energy of a free particle and the divergent integral
is regularized by a cutoff $\Lambda\simeq \pi/\bar{a}$, which is determined by the
mean scattering length $\bar{a}$.   

\section{UNIVERSALITY AND ZERO TEMPERATURE PHASE DIAGRAM}

In order to understand the nature of the critical points which give rise to universality in 
dilute, ultracold Fermi gases, it is useful to consider the onset transition
at zero temperature from the vacuum to the state with a finite density~\cite{sach11book}.
For attractive, two-component Fermi gases the zero density limit turns out to 
be fundamentally different, depending on whether the associated 
scattering length $a$ is positive or negative~\cite{niko07renorm}. For $a>0$, the existence of a two-body bound state 
with energy $\varepsilon_b=\hbar^2/ma^2$ implies that a finite density of Fermions appears - as a dilute gas of dimers -
already for a negative (Fermion) chemical potential $\mu>-\varepsilon_b/2$. Since the effective interaction 
between two dimers is repulsive with scattering length $a_{\rm dd}=0.6\, a$~\cite{petr04dimers},
one obtains a stable, dilute BEC.  
The line where $\mu(a)=-\varepsilon_b/2$ is thus a line of quantum critical points.
It separates the formally incompressible vacuum state from a weakly interacting  Bose superfluid. According to the Gross-Pitaevskii equation, its density vanishes linearly as $\mu +\varepsilon_b/2\to 0^+$.\\

For negative values of $a$, there is no bound state. A finite  density of Fermions thus only
appears for $\mu>0$, with $n(\mu, T=0)\sim\mu^{3/2}$ to leading order. Due to the attractive 
interaction, there is a pairing instability and the ground state is again a superfluid. 
In the low density regime $\mu\ll\hbar^2/ma^2$, which implies $k_F|a|\ll 1$, 
the superfluid is of the BCS-type. Pairing only affects an exponentially small
region of order $\exp{-\pi/(2k_F|a|)}$  around the Fermi surface. 
For negative values of the scattering length there is thus again 
a line of quantum critical points, now at $\mu=0$. It separates the vacuum 
from a weak coupling fermionic superfluid.  The situation is fundamentally different, however,
if the scattering length is infinite. There, lowering the density or chemical potential from 
a finite value towards zero, one never reaches a dilute gas of either Bosons or Fermions.  Instead,
the problem remains a strong coupling one for arbitrary small values of the density. \\

\begin{figure}[t]
\centering
\includegraphics[width=.9\columnwidth]{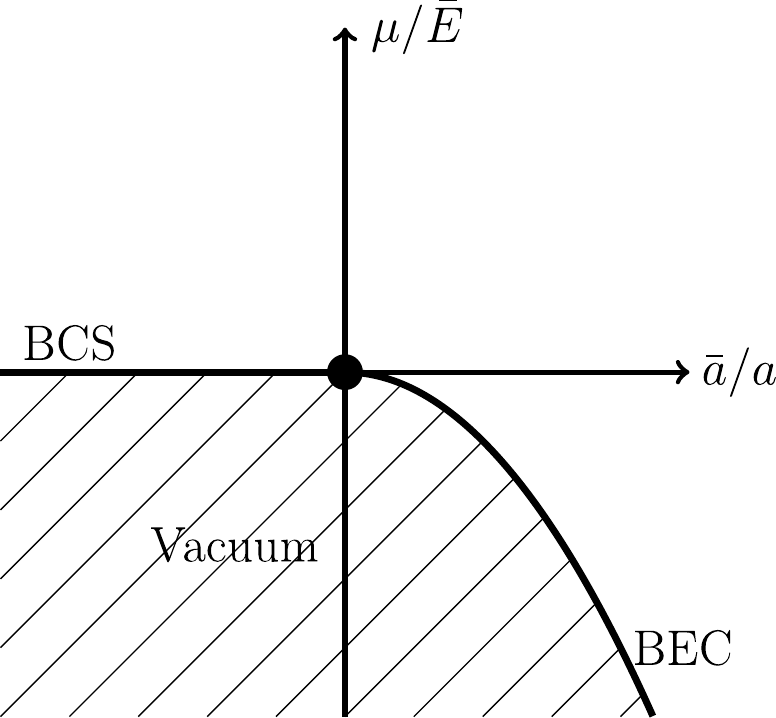}
\caption{Zero temperature phase diagram of a dilute Fermi gas with
attractive interactions. The unstable fixed point at zero chemical potential 
and infinite scattering length $\bar{a}/a=0$ describes the physics near a Feshbach resonance (from~\cite{niko07renorm}).} 
\label{fig:phase}
\end{figure}

The resulting zero temperature phase diagram for a spin-balanced gas is shown in Fig.~\ref{fig:phase}.
It contains two lines of continuous quantum phase transitions:
$\mu=0$ for negative and $\mu/\bar{E}=-(\bar{a}/a)^2$ for positive values of the scattering length.
The point at $1/a=0$ is the joint endpoint of these two lines and is thus a multicritical point.
As shown by Nikoli\'{c} and Sachdev~\cite{niko07renorm}, it is an unstable fixed point which has only three relevant perturbations. 
They are the dimensionless detuning 
$\nu=-\bar{a}/a$ away from resonance, the chemical potential $\mu$ and 
a possible finite difference $h=(\mu_{\shortuparrow}-\mu_{\shortdownarrow})/2$ of the 
chemical potentials for the two spin species.
A convenient representation for the scaling functions around this fixed point is obtained within a grand 
canonical description, where both the total particle number $\hat{N}=\hat{N}_{\shortuparrow}+\hat{N}_{\shortdownarrow}$ 
and the 'polarization' $\hat{S}=\hat{N}_{\shortuparrow}-\hat{N}_{\shortdownarrow}$ are controlled by
their conjugate thermodynamic variables $\mu$ and $h$. The associated partition function   
\begin{equation}
Z_V\left(\beta,\mu, h, 1/a\right)={\rm Tr}\,e^{-\beta(\hat{H}-\mu\hat{N}-h\hat{S})}=\exp{\left(V\beta p\right)}
\label{eq:grand-canonical}
\end{equation}
then determines the pressure $p$, from which the total density $n=n_{\shortuparrow}+n_{\shortdownarrow}$ 
and the spin density $s=n_{\shortuparrow}-n_{\shortdownarrow}$ follow by differentiation  
$n=\partial p/\partial\mu$ and $s=\partial p/\partial h$.  Its dependence on the three relevant scaling 
variables $\mu, \, h$ and $1/a$ can be expressed in terms of a universal function $f_p(x,y,z)$. 
A convenient definition is obtained by factorizing out the pressure $p^{(0)}(T,\mu)$ of a balanced two-component,
non-interacting Fermi gas in the form
 \begin{equation}
p(T,\mu,h,1/a)=p^{(0)}(T,\mu)\cdot f_p\,(\beta\mu, \beta h, \frac{\lambda_T}{a})
\label{eq:scaling1}
\end{equation}
where $\lambda_T=\hbar \sqrt{2\pi/ mT}$ is the thermal wavelength (we use units for the temperature where $k_B=1$). 
For the special case of a balanced, unitary gas, the function $f_p(\beta\mu, 0,0)$ has been determined 
experimentally from an integration of the directly measured density $n(\beta,\mu)$ in the relevant range 
between the non-degenerate limit  at $\beta\mu\simeq - 1.6$ down to~\cite{ku12normal} and also below the superfluid transition 
at $(\beta\mu)_c\simeq 2.5$~\cite{ku12superfluid}.  Its extension to finite values of $h$ will be calculated in section \ref{sec:LW} below, 
using the Luttinger-Ward formalism in the normal fluid regime. From  Eq.~(\ref{eq:scaling1}) all other thermodynamic 
properties can be deduced by differentiation. For example, the finite polarization due to
a nonzero field $h$ follows from 
\begin{equation}
s(T,\mu, h,1/a)=\beta p^{(0)}(T,\mu)\cdot \frac{\partial f_p\,(x, y,z)}{\partial y} \, .
\label{eq:spin-density}
\end{equation}

At zero temperature, the scaling function depends on two variables only, which is reduced
further to the single variable $h/\mu$ right at unitarity. For finite values of the scattering length,
both positive or negative, a convenient choice for the two scaling variables are the dimensionless 
chemical potential  $\bar{\mu}=\mu/(\varepsilon_b/2)$ and the dimensionless Zeeman field $\bar{h}=h/(\varepsilon_b/2)$. 
Factorizing out $p^{(0)}(0,\mu)\sim \mu (2m\mu)^{3/2}/\hbar^3$, the scaling functions at zero temperature may then be defined by
 \begin{equation}
\frac{p(\mu,h,1/a)}{p^{(0)}(0, \mu)}=f_{\pm}\,(\bar{\mu}, \bar{h}) \,\to\,  f(h/\mu)\; {\rm at} \; 1/a=0\, ,
\label{eq:scaling2}
\end{equation}
where the subscript $\pm$ differentiates between positive or negative values of $1/a$. In particular, the 
value $f(0)=\xi_s^{-3/2}\simeq 4.44$ of the scaling function, which characterizes the ground state of the balanced 
Fermi gas at unitarity, is fixed by the Bertsch parameter $\xi_s\simeq 0.37$. This theoretical value is predicted by 
a Pad\'e resummation of the $\varepsilon=4-d$ expansion for the unitary gas developed by Nishida and Son,
which gives $\xi_s=0.365\pm 0.01$~\cite{nish09eps,nish12book}. It agrees perfectly 
both with the measured value $\xi_s=0.37\pm 0.01$, taking into account the 
precise position $B_0=832.18\,$G of the $^6$Li resonance~\cite{zuer13Li-resonance} and also
with the result $\xi_s=0.36$ obtained from a diagrammatic calculation based on the Luttinger-Ward
approach \cite{haus07bcsbec}. \\  

Equations~(\ref{eq:scaling1}) and~(\ref{eq:scaling2}) generalize results which have been 
formulated for the special case of the unitary gas at finite temperature by Ho~\cite{ho04uni} 
and at zero temperature in Refs.~\cite{chev06polaron, bulg07asym}.
The proposed scaling is based on the argument that the ratio $p/p^{(0)}$ can only depend 
on the dimensionless  variables $\beta\mu$, $\beta h$ and $\lambda_T/a$, provided $a$ is the 
single relevant length scale which fully characterizes the interaction
\footnote{Dimensional analysis is sufficient to formulate a scaling function for the pressure because it does not  develop 
an anomalous dimension, in contrast to observables like the collisional relaxation rate~\cite{petr05dimer} or the
closed channel fraction~(\ref{eq:Nb}), which involve the additional microscopic lengths $\bar{a}$ or $r^\star$.}.
The insight that~(\ref{eq:scaling1}) and~(\ref{eq:scaling2}) 
are universal scaling functions associated with the unstable fixed point at zero density 
discussed above provides a lot of additional information, however.\\
\begin{strip}
\centering
\includegraphics[width=\textwidth]{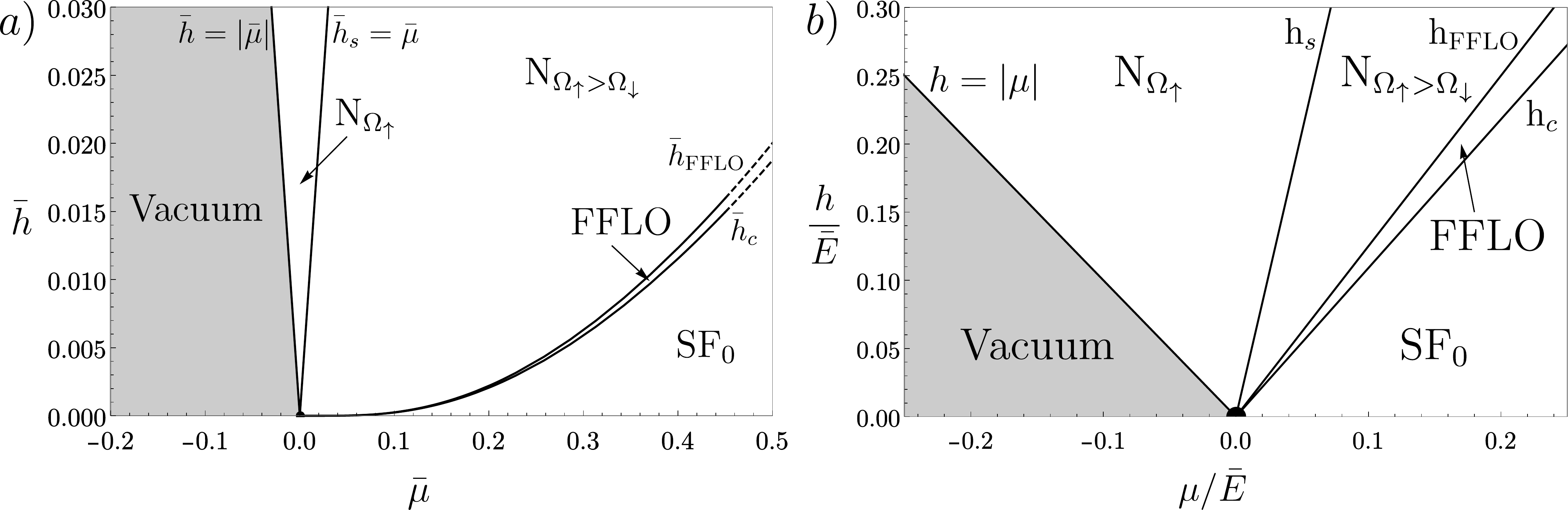}
\captionof{figure}{Panel~a): Universal phase diagram for negative values of $1/a$ with the finite density phases: fully polarized gas $N_{\Omega_\shortuparrow}$, normal fluid with finite spin population of the minority atoms $N_{\Omega_\shortuparrow>\Omega_\shortdownarrow}$ and the superfluid FFLO and balanced BCS phase, the latter denoted as $\text{SF}_\text{0}$. The dashed lines indicate the crossover of the critical fields towards the strong coupling regime. Panel~b): Unitary limit of the universal phase diagram, where all phase boundaries extend linearly from the quantum multicritical point. Note that the slopes of $h_c$ and $h_{\text{FFLO}} $ represent the Luttinger-Ward results.}
\label{fig:phase_a<0}
\end{strip}
\noindent In particular, focusing on zero temperature,  
the fact that the crossovers between the strong coupling fixed point at $\mu=1/a=h=0$ and the two
possible weak coupling fixed points at finite values of the scattering length occur at $\bar{\mu}=\mathcal{O}(1)$ and
$\bar{h}=\mathcal{O}(1)$ fully determines the scaling function $f_{\pm}\,(\bar{\mu}, \bar{h})$ in both
limits $ \bar{\mu},\bar{h}\ll 1$ and $ \bar{\mu},\bar{h}\gg 1$ (see below). As a result,
the phase diagram at \emph{any} finite value of the scattering  length can be collapsed to just a single one on the BCS
or the BEC side by a simple rescaling of $\mu$ and $h$. \\

For the phase diagram at zero temperature but finite values of the field $h$, the unitary case 
and the one for negative scattering length 
(both depicted in Fig.~\ref{fig:phase_a<0})
turn out to be the simplest conceptually. Starting with 
the unitary gas, the vacuum state at $\mu<0$ evolves into a fully polarized Fermi gas at $h=|\mu|$,
because, for negative values of  the chemical potential, $h>|\mu|$ implies $\mu_\shortuparrow>0$. 
The same state can also be reached for $\mu>0$, provided $h>h_s$ is larger than a 'saturation field'
$h_s$ beyond which the Fermi gas is fully polarized. As realized by Chevy~\cite{chev06polaron}, this field is 
determined by calculating the addition energy of a single down spin added to a Fermi sea of up spins. 
A precise solution of the associated Fermi polaron problem shows that the ratio 
$\eta=\mu_\shortdownarrow/\mu_\shortuparrow=-0.615$ of the chemical potentials has a universal 
value at unitarity~\cite{prok08polaron}.  Due to
\begin{equation}
\label{eq:eta-definition}
\eta=\frac{\mu_\shortdownarrow}{\mu_\shortuparrow}=\frac{1-h/\mu}{1+h/\mu}\leq 1
\end{equation}
this ratio fixes the corresponding saturation field $(h/\mu)_s=(1-\eta)/(1+\eta)=4.19$. Starting 
from small values of the field $h$, the gapped superfluid at $h=0$ acquires a finite polarization $s\ne 0$
beyond a Clogston-Chandrasekhar (CC) field $h_c$~\cite{zwie05imbalance,part06phase}.  
From our Luttinger-Ward calculation below, this field is at 
$(h/\mu)_c=1.09\pm 0.05$. A discussion of how this result compares with previous theoretical and also 
experimental values will be given in section \ref{sec:LW}. A crucial and still open issue is the precise  
nature of the ground state in the regime $h_c<h<h_s$. The standard assumption is that it is 
a normal Fermi liquid with two Fermi surfaces. Their volumes $\Omega_{\shortuparrow,\shortdownarrow}$
in momentum space are connected with the finite polarization $s\ne 0$ by the exact relation
\begin{equation}
\label{eq:FS-volume}
s=\frac{\Omega_{\shortuparrow}-\Omega_{\shortdownarrow}}{\left(2\pi\right)^3}
\end{equation} 
derived, for homogeneous phases, by Sachdev and Yang~\cite{sach06lutt}. The relation shows, that any phase with 
a finite polarization must have at least one Fermi surface. In phases which break translation invariance like
FFLO, the relation holds only modulo the volume of the unit cell in the corresponding reciprocal lattice. 
Our results below indicate that, in fact,  an intermediate FFLO phase appears beyond the 
CC - limit even for the unitary gas. Specifically, 
we find that the vertex associated with a pairing instability diverges at a 
finite rather than at zero center of mass momentum $\mathbf{Q}$ at the lowest accessible temperatures in the
regime immediately above $h_c$. The conclusion is also supported by a continuity argument: as will be
discussed below, there is an FFLO line $h_{\rm FFLO}(\bar{\mu})$ in the universal phase diagram 
covering negative values of the scattering length. In the limit $\varepsilon_b\to 0$ or 
$\bar{\mu}\gg 1$, this line must continuously evolve into a line of phase transitions for the gas near unitarity.\\
\begin{strip}
\centering
\includegraphics[width=\textwidth]{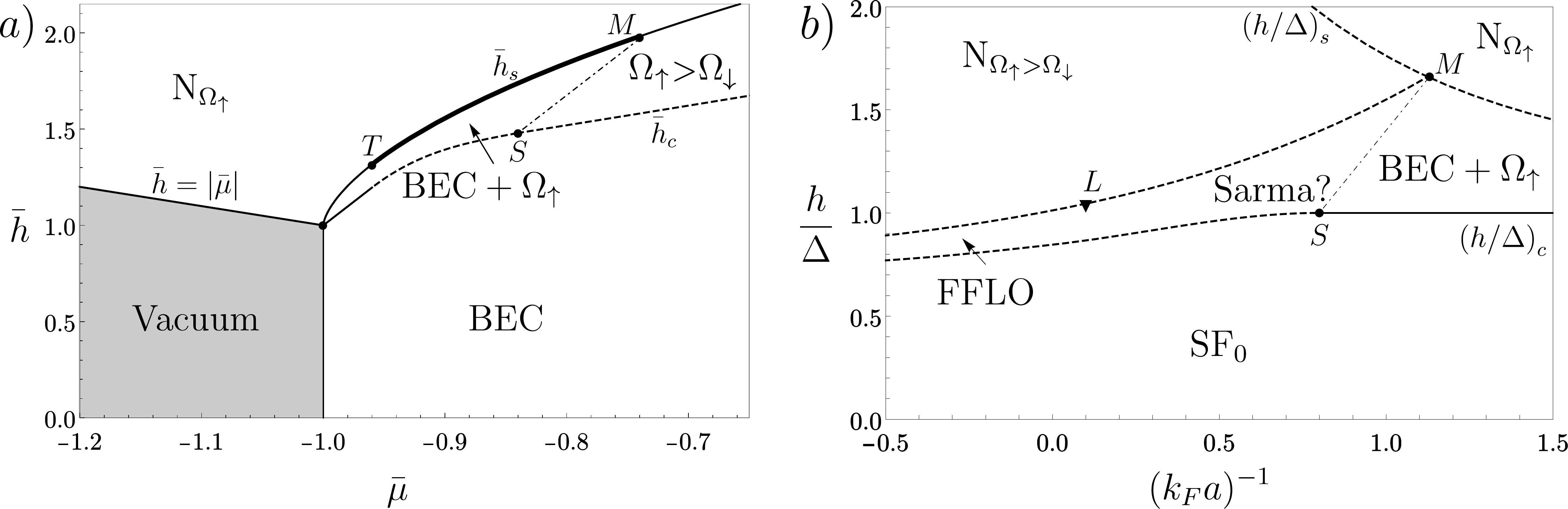}
\captionof{figure}{Panel~a): Universal phase diagram on the BEC side. BEC+$\Omega_\shortuparrow$ is a condensate with an
additional Fermi surface. Beyond the point $S$, an inhomogeneous superfluid with two Fermi surfaces appears just above $\bar{h}_c$. The thick solid line between $T$ and $M$ denotes a first order transition. Along the dash-dotted line $SM$ $\Omega_\shortdownarrow=0$. Panel~b): A more detailed version of the phase diagram due to Son and Stephanov, including both an FFLO and a Sarma phase. The Lifshitz point $L$ (omitted in Panel~a)), where FFLO disappears, is close to unitarity.}
\label{fig:phase_a>0}
\end{strip}
\noindent Unless the nature of the transition changes at some finite value $\bar{\mu}=\mathcal{O}(1)$ along
this line
\footnote{This happens in mean field calculations, where FFLO disappears in the ground state 
at a finite negative $1/(k_Fa)^*= -2.86$~\cite{pari07finitetemp} or $1/(k_Fa)^*=-0.46$~\cite{shee07phase}.},
an inhomogeneous superfluid with FFLO order must be present at unitarity. From our numerical 
results below, we estimate a critical value $(h/\mu)_{\rm FFLO}=1.28\pm 0.15$. The ground state beyond 
$h_{\rm FFLO}$, which again has  two Fermi surfaces, is likely to be unstable to p-wave pairing around the majority 
Fermi surface as discussed by Kagan and Chubukov for imbalanced Fermi gases with repulsive interactions in the regime $k_F a \ll 1$~\cite{kaga89pwave}, see also~\cite{bara96pwave}. More recent calculations by Patton and Sheehy for the unitary gas indicate that the transition temperatures can reach up to $0.03\,\varepsilon_{F\shortuparrow}$~\cite{patt11}. 
For the following discussion, we will neglect this and assume it is a Fermi liquid state, as indicated 
experimentally~\cite{nasc11fermiliquid}. The resulting phase diagram is shown in Figure \ref{fig:phase_a<0}\,b). Apart from
the vacuum transition line $h=|\mu|$, it features three straight lines which, as a consequence of scale invariance,
are perfectly linear up to the cutoff energy $\bar{E}$. Their slopes are universal numbers,
which are just the values of $h/\mu$ where the scaling function $f(h/\mu)$ defined in~(\ref{eq:scaling2}) exhibits singularities. 
In particular, at the first order CC - transition, 
$f(h/\mu)$ has a discontinuity in slope from zero below $(h/\mu)_c$ to a positive value above.\\

The $h$ versus $\mu$ phase diagram at negative values of the scattering length is determined by the 
fact that the limit $\mu\ll \hbar^2/ma^2$ corresponds to a Fermi superfluid of the BCS-type, with a crossover to
the strong coupling fixed point for $\bar{\mu}=\mu/(\varepsilon_b/2)=\mathcal{O}(1)$. Indeed, the line $\mu=0$ at finite negative
values of $1/a$ is a line of quantum critical points which all share the same fixed point, namely a weak 
coupling BCS superfluid. In the limit $\bar{\mu}\ll 1$, the boundary of the balanced superfluid phase is 
thus given by the standard result $h_c=\Delta/\sqrt{2}$~\cite{chan62,clog62}, which yields
\begin{equation}
\label{eq:CC-line}
\bar{h}_c(\bar{\mu}\ll 1)=\bar{\mu}\cdot\frac{(2/e)^{7/3}}{\sqrt{2}}\,\exp{(-\pi/2\sqrt{\bar{\mu}})}\,,
\end{equation} 
where we have used the exact expression for the gap $\Delta$~\cite{gork61}. This line extends 
in a continuous manner to the strong coupling regime $\bar{\mu}\gg 1$, where $(h/\mu)_c \simeq 1.09$ 
approaches the strictly linear behavior at the unitary fixed point. Note that the limit $\bar{\mu}\gg 1$ is
compatible with the condition $\mu\ll\bar{E}$, i.e. the domain of applicability of the contact interaction 
Hamiltonian~(\ref{eq:H-direct}), because, as  discussed above, for open channel dominated resonances, 
a wide range of scattering lengths $|a|\gg\bar{a}$ is accessible. A similar crossover between the
weak coupling and the unitary gas fixed point appears for the critical temperature, where 
$(T/\mu)_c=0.277\,\exp{(-\pi/2\sqrt{\bar{\mu}})}$ for $\bar{\mu}\ll 1$ evolves continuously into 
$(T/\mu)_c\simeq 0.4$ for $\bar{\mu}\gg 1$. The latter dependence characterizes the gas at unitarity, 
where $T_c$ scales linearly with $\mu$~\cite{enss12QCP}. 
The fact that the CC - line extends in a continuous manner up to unitarity and even beyond
to positive scattering lengths is connected with the fact that the so called splitting point, where - coming from
the BEC side at positive values of $1/a$ - the continuous transition from a balanced to a polarized superfluid 
with a finite Fermi surface of majority atoms splits into two first order transitions~\cite{son06imbalance} - is on 
the BEC side at $1/a>0$. This will be discussed further below. Beyond the CC - transition, a weak coupling 
Fermi gas turns into an inhomogeneous superfluid with an order parameter which varies periodically like 
$\cos{(\mathbf{Q}\cdot\mathbf{x})}$. This phase, which is denoted generically by FFLO in the 
following, is the ground state in a small window $1/\sqrt{2}<h/\Delta<0.754$~\cite{LO64,FF64}. 
It defines an FFLO line $h_{\rm FFLO}(\bar{\mu}\ll 1)= 0.754\,\Delta$ just above the CC - line~(\ref{eq:CC-line}) 
with the same dependence on $\bar{\mu}$. For $\bar{\mu}\gg 1$, this line again acquires the linear 
dependence $(h/\mu)_{\rm FFLO}\simeq1.28$ characteristic for the behavior at unitarity.   
Similarly, the saturation field $h_s$ beyond which only a single Fermi surface remains, has a simple
form  $h_s(\bar{\mu}\ll 1)=\mu$ in the weak coupling limit, because pairing is suppressed exponentially.
Thus, to leading order, the saturation field is given by its value in a non-interacting Fermi gas.
For $\bar{\mu}\gg 1$, in turn,  the dependence is again linear, but the slope is now given by the universal number
$4.19$ which characterizes the strong coupling fixed point discussed above. The complete phase diagram is shown in Figure~\ref{fig:phase_a<0}a),
restricted to the regime $\bar{\mu}\ll 1$ where everything is covered by standard results for pairing of 
Fermions in a weak coupling approximation. The new aspect associated with the crossover between the 
 weak and strong coupling fixed points discussed above is that, using the proper dimensionless units
 $\bar{\mu}$ and $\bar{h}$, this diagram is \emph{the} universal phase diagram for all negative 
 values of the scattering length.  Arbitrary values of $1/a<0$ can thus be collapsed into a single
 diagram whose behavior is known quantitatively in both the weak and strong coupling limit  $\bar{\mu}\ll 1$ 
 or $\bar{\mu}\gg 1$.  
 Note also, that the $h$ versus $\mu$ diagram is directly accessible in the presence of a harmonic trap.
 Indeed, within a local density approximation, the chemical potential $\mu(r)=\mu(0)-m\omega^2r^2/2$ 
 decreases quadratically with the distance $r$ from the center of the trap, 
 while $h=(\mu_\shortuparrow-\mu_\shortdownarrow)/2$ stays constant. The different phases thus 
 appear according to a horizontal line from right to left in Figure 2. \\  
 
 Finally, we turn to the most complex case of the phase diagram at finite positive values of $1/a$, 
 where the ground state at $h=0$ is a dilute, repulsive BEC with a dimer-dimer scattering length 
 $a_{dd}=0.6a$. At $h=0$, a molecular BEC appears for $\mu>-\varepsilon_b/2$. At finite values of $h$, 
 however,  even smaller values of the chemical potential are physically relevant.
 Indeed, the transition out of the vacuum state at $\mu<-\varepsilon_b/2$ again appears at $h=|\mu|$ into
 a fully polarized gas of $\shortuparrow$-Fermions. The boundary line of this phase defines the
 associated saturation field $h_s$. As in the case of negative scattering lengths, this line
 is fully determined by considering the problem of adding a single down spin to an up-spin Fermi sea.
 Close to the vacuum point at $\mu=-\varepsilon_b/2$ and $h=\varepsilon_b/2$, this problem exhibits a 
 bound state rather than a stable Fermi polaron~\cite{prok08polaron}. Over a remarkably wide range of coupling constants,
the chemical potential for adding a single down spin on this molecular branch obeys
$\mu_\shortdownarrow=-\varepsilon_b - \mu_{\shortuparrow}+g_{ad}n_\shortuparrow$~\cite{prok08polaron,punk09molaron}.  
Here, $n_\uparrow$ is the density of the noninteracting majority atoms while $g_{ad}= 3\pi \hbar^2 a_{ad}/m$ describes the atom-dimer 
repulsion associated with a corresponding scattering length $a_{ad}=1.18 a$ \cite{petr05dimer}.
The resulting ratio $\eta=\mu_\shortdownarrow/\mu_\shortuparrow$ then determines the saturation field
 \begin{equation}
\label{eq:h_s-line}
h_s=\frac{\varepsilon_b}{2}+{\rm const}\left(\frac{2\mu+\varepsilon_b}{g_{ad}}\right)^{2/3}+\ldots\,,
\end{equation}
where $\text{const}=\hbar^2(6\pi^2)^{2/3}/(2m)$ can be directly obtained from the ground state energy of the majority atoms.
This relation remains quantitatively valid for all values of $\bar{\mu}$ shown in Fig.~\ref{fig:phase_a>0}\,a)~\cite{prok08polaron,punk09molaron}.\\
To determine the CC - field for positive values of the scattering length,  we use the fact that on the bosonic side of the splitting point $S$, $h_c\equiv\Delta$ 
is given exactly by the excitation energy gap $\Delta$ for Fermi quasiparticles~\cite{son06imbalance} (see below). 
Inserting the known asymptotic form of $\Delta$ in the BEC-limit~\cite{gior08review},  the critical field strength
\begin{align}
\begin{split}
\label{eq:h_c-line}
h_c &=\frac{\varepsilon_b}{2}+\left(3\,\frac{a_{ad}}{a_{dd}}-1\right)\left(\mu+\varepsilon_b/2\right)+\ldots
\end{split}
\end{align} 
exhibits a linear dependence on $\mu+\varepsilon_b/2$ from its onset out of the vacuum, with a universal slope $4.9$.\\

As shown by Son and Stephanov~\cite{son06imbalance}, the nature of the CC-line changes
when the quasiparticle spectrum ceases to have its minimum at a finite value 
$k_0$ of the momentum. Beyond this point, majority atoms may form a Fermi surface
with volume $\Omega_\shortuparrow=(2\pi)^3\, s$ on top of a BEC of pairs.
The critical field $h_c\equiv\Delta$ is then identical with the excitation gap at 
$\mathbf{k}=0$ (note that this gap is now different from the gap parameter
defined by the short distance limit of the anomalous Green function
$\mathcal{F}(\mathbf{x}\to 0, \tau=0^-)$). Within a mean field description, 
the quasiparticle dispersion has its minimum at $\mathbf{k}=0$
in the regime where the chemical potential is negative. From a detailed
calculation of the spectrum within a Luttinger-Ward approach, however, 
it turns out that the associated critical value of the parameter $v=1/k_Fa$ 
is $v_S\simeq 0.8$~\cite{haus09rf}, which is about a factor of two larger than 
the point where $\mu$ changes sign. An equivalent of the point $S$ has been
determined from purely thermodynamic considerations by Pilati and Giorgini~\cite{pila08phase},
who estimate the regime for which the balanced superfluid phase cannot be
polarized in a continuous manner as $v\leq 0.53$. This and also the 
corresponding result $v\leq 0.66$ given in~\cite{chev10review},
is considerably smaller than the value obtained from an explicit calculation
of the quasiparticle spectrum. The origin of this discrepancy remains open. 
Using the Luttinger-Ward results for the splitting point~\cite{haus09rf} at $v_S=0.8$, $\Delta(v_S)=0.95 \varepsilon_F $ and
$\mu(v_S)=-0.54 \varepsilon_F$,  its coordinates in Fig.~\ref{fig:phase_a>0}\,a) are found to be 
$(\bar{\mu},\bar{h})=(-0.84, 1.48)$ which is already outside of the regime where the expansion for $h_c$ around the onset from the vacuum from equation \eqref{eq:h_c-line} holds.
Furthermore, as argued by Son and Stephanov~\cite{son06imbalance}, there exists a Lifshitz point $L$ along the line
which extends the boundary $h_{\rm FFLO}$ of the FFLO phase at weak coupling into the strong coupling regime near unitarity. 
From our results in section \ref{sec:LW}, this point is rather close to $1/a=0$. In the following, we assume it is still at positive scattering 
lengths. As a result, as shown in Figure \ref{fig:phase_a>0}\,b), the FFLO phase of the unitary gas lies between the balanced superfluid SF$_0$ and a normal phase
with two Fermi surfaces, rather than a further unconventional superfluid.  \\   

Along the saturation line $h_s$ one finds two special points:
The polaron to molecule transition point $M$, whose position is known quite reliably both from bold diagrammatic Monte Carlo~\cite{prok08polaron} and 
from a molecular ansatz~\cite{punk09molaron}. It is located at $1/(k_{F\shortuparrow} a)=0.9$ and $\eta_\text{M}=-2.2$.
For a finite density of minority atoms, this point is very likely the endpoint of the line, where the volume $\Omega_\shortdownarrow$ 
of the  minority Fermi surface vanishes. The corresponding coordinates in Fig \ref{fig:phase_a>0} are $(\bar{\mu},\bar{h})_\text{M}=(-0.74,1.97)$.
In a diagram $(v,h/\Delta)$, which corresponds to a fixed total density $n$ and thus an associated Fermi wave vector $k_F=(3\pi^2 n)^{1/3}$,
the point $M$ is at $(v,h/\Delta)_\text{M}=(1.13, 1.66)$. This compares very well with our result $(v,h/\Delta)_\text{M}\simeq(1.13, 1.65)$, obtained from the position on the critical line, where the polarization reaches unity according to our self-consistent Luttinger-Ward approach described below. Closer to the BEC limit, there is also a tricritical point $T$, beyond which the mixture
of a molecular BEC and a Fermi gas of majority atoms exhibit phase separation. The transition 
to a fully polarized gas is then of first order~\cite{pari07finitetemp}. From variational Monte Carlo calculations, the point $T$ is located 
at $1/(k_{F\shortuparrow} a)=1.7$~\cite{pila08phase}, which translates to $(\bar{\mu}, \bar{h})_\text{T}=(-0.96, 1.31)$.\\

Following the qualitative but rather general description of the phase diagram by
Son and Stephanov~\cite{son06imbalance}, we assume that one of the two lines
entering the splitting point $S$ from the BEC side defines a line, at which the volume
$\Omega_\shortdownarrow$ of the minority Fermi surface vanishes, ending at the
polaron to molecule transition point $M$. In this simplest possible form of the zero temperature phase diagram which is consistent with all
known data, the unitary gas necessarily exhibits an FFLO phase. Moreover, there is a superfluid with two Fermi surfaces 
in the triangle LMS, which is likely to be a homogeneous Sarma phase. Whether the latter is indeed an absolute minimum of
the grand canonical potential or possibly only a metastable phase is open.

\section{LUTTINGER-WARD THEORY AT FINITE SPIN-IMBALANCE}\label{sec:LW}

To study the thermodynamic properties in the presence of unequal spin densities,
we use a diagrammatic approach based on Green functions in imaginary time $\tau \in [0, \beta]$. 
Following the standard notation~\cite{abri75}, it is defined as
\begin{align}
\mathcal{G}_\sigma \left(\mathbf x, \tau \right) = - \left \langle \mathcal{T} \left[ \hat\Psi_\sigma \left(\mathbf x, \tau \right) \hat{\Psi}_\sigma^\dagger\left(\mathbf 0 , 0\right) \right] \right \rangle\, .
\end{align}
The index $\sigma= \shortuparrow, \shortdownarrow$ denotes the spin orientation and $\mathcal{T}$ is the time-ordering operator. 
Since we only consider the normal phase in our present work, no anomalous expectation values appear. The densities can be extracted via the standard relation
\begin{align}
n_\sigma = \mathcal{G}_\sigma\left( \mathbf x = \mathbf 0, \tau = 0^- \right)\, .
\label{eq:density}
\end{align}
In the following, we restrict ourselves to the dominant instability of the Fermi gas with attractive contact interactions, which is s-wave pairing in a relative singlet configuration. This
allows to restrict the momentum dependence of the vertex function $\Gamma$ to a single momentum and frequency. 
Furthermore, in the zero range limit, $\Gamma$ becomes a scalar in spin space, which contains only the center of mass dynamics of the pairs  
\begin{align}
\begin{split}
\Gamma(\mathbf{x},\tau)=&\bar{g}(\Lambda)\,\delta(\tau)\delta(\mathbf{x})\\&-\bar{g}^2(\Lambda)\,
\langle  \mathcal{T}\, \left(\hat{\psi}_{\shortdownarrow}\hat{\psi}_{\shortuparrow}\right)(\mathbf{x},\tau)\,\left(\hat{\psi}_{\shortuparrow}^\dagger\hat{\psi}_{\shortdownarrow}^\dagger\right)
(\bm{0},0)\rangle\, .
\label{eq:Gamma-def}
\end{split}
\end{align}
Its short distance and short time limit determines the Tan contact density
\begin{align}
\begin{split}
\frac{\hbar^4}{m^2}\mathcal{C}&=
-\Gamma(\mathbf{x}=\bm{0},\tau=0^-)\\ &= \lim_{\Lambda\to\infty}\,\bar{g}^2(\Lambda)\,\langle \hat{\psi}_{\shortuparrow}^{\dagger}(\mathbf{x})
\hat{\psi}_{\shortdownarrow}^{\dagger}(\mathbf{x})
 \hat{\psi}_{\shortdownarrow}(\mathbf{x})\hat{\psi}_{\shortuparrow}(\mathbf{x})\rangle \, ,
 \label{eq:contact-Delta}
 \end{split}
 \end{align}
which, above the superfluid transition, has no anomalous contribution involving the gap para\-meter $\Delta$~\cite{haus09rf,zwer14varenna}.\\

The grand potential $\Omega(T,\mu, h, 1/a, V)= - p\, V$ which directly determines the pressure $p$, 
can be represented in terms of the formally exact Luttinger-Ward  functional~\cite{lutt60}  
\begin{equation}
\Omega[G] = \beta^{-1} \bigl(
\text{Tr} \{ \ln[G] + [1-G_{0}^{-1} G] \} + \Phi[G] \bigr ) \,,
\label{eq:LW-functional}
\end{equation}
where  $G=\{\mathcal G_{\shortuparrow}, \mathcal G_{\shortdownarrow}\}$ denotes the two independent components of the interacting Green function and $G_0=\{\mathcal G_{\shortuparrow,0}, \mathcal G_{\shortdownarrow,0}\}$ contains the bare Green functions
\begin{align}
\mathcal{G}_{\sigma,0}(\mathbf k, \omega_n) = \frac{1}{i \hbar\,\omega_n - \left(\varepsilon_k-\mu_\sigma\right)}\, .
\end{align}
They involve fermionic Matsubara frequencies $\hbar\omega_n = 2\pi(n+1/2)\, T$ and the 
free particle dispersion relation $\varepsilon_k = \hbar^2 k^2 / (2 m)$. The nontrivial part $\Phi$ of the
Luttinger-Ward  functional contains all topologically allowed skeleton Feynman diagrams. It is
defined by the property that the physical, interacting Green function solves the self-consistent Dyson equation
\begin{equation}
\mathcal G _{\sigma}^{-1}=\mathcal G_{\sigma,0}^{-1}-\Sigma_\sigma\,[G]\, ,
\label{eq:Dyson}
\end{equation} 
where the spin-dependent self-energy is given by the functional derivative
\begin{equation}
\Sigma_\sigma[G]=\frac{\delta\Phi[G]}{\delta \mathcal{G}_\sigma}\,.
\label{eq:sigma1} 
\end{equation}
The conserving nature of the formalism guarantees that all thermodynamic relations are obeyed, irrespective of the approximations used for $\Phi$, 
provided that the Green function $G$ is a self-consistent solution to the equations \eqref{eq:Dyson} and \eqref{eq:sigma1}~\cite{baym61}. 
In our self-consistent theory for the imbalanced Fermi gas, we constrain $\Phi$ to the particle-particle ladder.
This approximation captures the BCS instability, however it does not fully account for particle-hole fluctuations. In the weak coupling limit,
the latter have been included properly for a zero range interaction by Gorkov and Melik-Barkhudarov~\cite{gork61}, giving rise to a reduction of 
the BCS critical temperature by a factor of $(4e)^{1/3}\simeq 2.22$.
In the relevant regime around unitarity, these fluctuations are apparently strongly suppressed. Indeed, for a balanced Fermi gas at $h=0$, 
both the critical temperature $T_c/T_F\simeq 0.16$ and the Bertsch parameter $\xi_s=0.36$ at $T=0$, obtained from a Luttinger-Ward approach 
based on including only the particle-particle ladder~\cite{haus07bcsbec},
agree very well with the corresponding experimental values~\cite{ku12superfluid}. 
Within this approximation, the vertex function can be written in the form
\begin{align}
\Gamma(\mathbf Q ,\Omega_n) =\frac{1}{\frac{1}{g} + M\left(\mathbf Q, \Omega_n\right) }\,,
\label{eq:Vertex-ladder}
\end{align}
which results from the geometric series of the particle-particle bubble diagram 
\begin{align}
\begin{split}
&M(\mathbf{Q}, \Omega_n) =\\ \!&\int{\!\!\frac{d^3 k}{\left(2 \pi \right)^3}\!\left[ \frac{1}{\beta} \!\sum_m {\mathcal G_\shortuparrow\!\left(\mathbf k , \omega_m \right) \mathcal G_\shortdownarrow \!\left(\mathbf Q - \mathbf k, \Omega_n\! - \omega_m\right)} - \frac{1}{2 \varepsilon_k} \right]}\,.
\label{eq:bubble}
\end{split}
\end{align}
Note that the last term renders the result finite and has been introduced by the renormalization scheme of equation \eqref{eq:gbar} as we have already taken the limit $\Lambda \to\infty$ in equation \eqref{eq:Vertex-ladder}. 
The self-energy is connected with the vertex and the exact Green function via
\begin{align}
\begin{split}
&\Sigma_\sigma(\mathbf k, \omega_n) =\\& \int{\frac{d^3 Q}{\left(2 \pi \right)^3} \frac{1}{\beta}\sum_{m}{ \Gamma \left(\mathbf Q , \Omega_m\right) \mathcal G_{\bar{\sigma}}\left(\mathbf Q - \mathbf k, \Omega_m - \omega_n\right)  }}\,,
\label{eq:sigma2}
\end{split}
\end{align}
where $\bar{\sigma}$ is the spin orientation complementary to $\sigma$.
Together with \eqref{eq:Dyson} and \eqref{eq:Vertex-ladder},  this gives a closed set of equations for the 
two independent Green functions $\mathcal G_\sigma\left(\mathbf k , \omega_n \right)$. \\

\begin{figure}[t]
\centering
\includegraphics[width=.95\columnwidth]{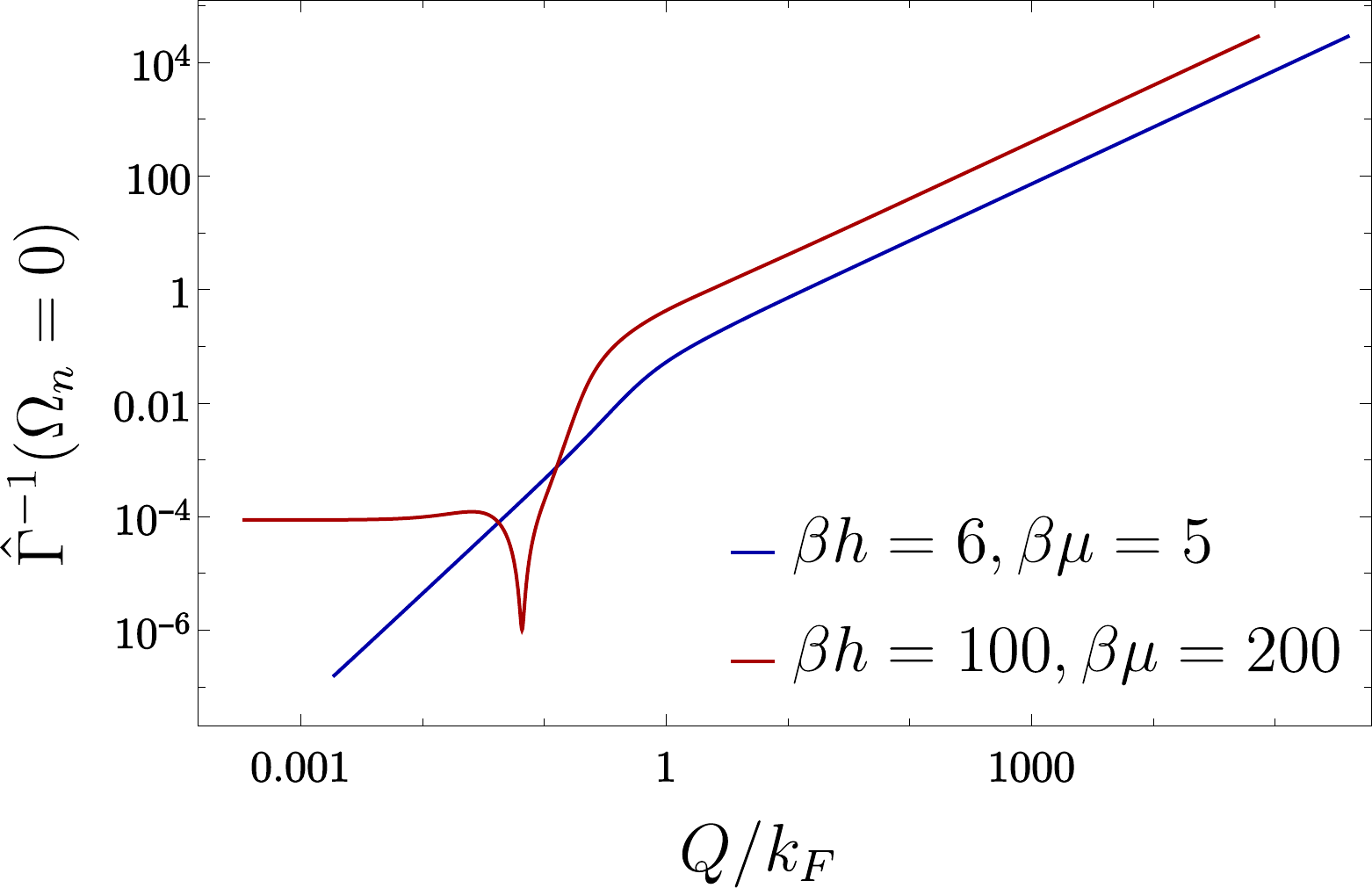}
\caption{(Color online) Double logarithmic plot of two typical critical inverse vertices. Blue line: At relatively high temperatures and positive scattering lengths pair formation at $Q=0$ causes a phase transition to a homogeneous superfluid phase. Criticality is ensured by demanding $\hat{\Gamma}^{-1}\sim Q^2$ at small momenta. Red line: At sufficiently low temperatures and attractive interactions the system becomes unstable towards pair formation at a finite center-of-mass momentum, resulting in a phase transition towards FFLO.} 
\label{fig:inverse_vertex}
\end{figure}

The equations \eqref{eq:Vertex-ladder} - \eqref{eq:sigma2} together with the Dyson equation \eqref{eq:Dyson} are solved numerically in an  
iterative manner,  until convergence for each set of initial, dimensionless parameters $\beta \mu, \beta h$ and $\lambda_T/a$ has been reached. 
The results may then be converted to the more common canonical description, 
where the momentum  and energy scales are expressed in terms of the Fermi wave vector $k_F=\left(3 \pi^2(n_\shortuparrow + n_\shortdownarrow)\right)^{1/3}$ and the 
corresponding Fermi energy $\varepsilon_F= \hbar^2 k_F^2/2 m$ of an ideal Fermi gas with the same total particle density. The individual spin densities 
$n_\shortuparrow, n_\shortdownarrow$ follow from \eqref{eq:density}.\\
For an efficient numerical evaluation of the convolutions in equations \eqref{eq:bubble} and 
\eqref{eq:sigma2}, we use the standard procedure of Fourier transforming the factors in the integrand to real space and imaginary time. These are then multiplied and converted back to momentum and frequency space by an inverse Fourier transformation. As both the Green and vertex function show algebraically decaying asymptotics in $k$ and $\omega_n$, we need an appropriately 
tailored version of the Fourier transformation. The 
transformation between $\tau$ and $\omega_n$ or $\Omega_n$, respectively, is performed by a spline Fourier transform \cite{haus07bcsbec}. For this method we first interpolate the function by a 
spline of up to fifth order on a grid of 512 appropriately spaced points in the $\tau$-interval and on the same number of exact Matsubara frequencies where we use a grid with exponential scaling in 
the high frequency regime to correctly capture the power-law tails, which are important for the computation of the thermodynamic quantities. The subsequent Fourier transformation of the spline yields a drastic suppression of discretization errors compared to a direct discrete Fourier transform. For the $\mathbf k \leftrightarrow \mathbf x$ 
transformations we typically use a logarithmic Fourier transform~\cite{lft88} based on a fast Fourier transform on a grid of 512 equidistant points on a logarithmic lattice.
Alternatively, for $\beta \mu \gg 1$, we switch to a spline Fourier transformation with adaptively increased grid density around the Fermi surface and the FFLO vector, which are both subject to continuous updates during the iteration. \\
The instability to the formation of a superfluid is signaled by a divergent vertex or equivalently by a zero of the inverse vertex. In the standard case of pairing with vanishing 
center of mass momentum, we locate the transition point in parameter space by the criterion that $\Gamma^{-1}\left(\mathbf Q \to \mathbf 0, \Omega_n=0 \right)\sim\mathbf{Q}^2$ 
approaches zero like the square of the center of mass momentum~\cite{abri75}. The quadratic behavior in $\mathbf{Q}$ is well resolved in our numerical data \footnote{Strictly speaking $\Gamma^{-1}\left(\mathbf Q, \Omega_n=0\right)\sim |\mathbf{Q}|^{2-\eta}$ involves an anomalous dimension $\eta$ which is, however, very small 
for the transition to superfluidity in three dimensions and is anyway not properly contained in our approach.} (see Figure \ref{fig:inverse_vertex}) 
while for very large values of $ |\mathbf{Q}|$ we obtain a linear behavior
$\Gamma^{-1}\left(\mathbf Q, \Omega_n=0\right)\sim |\mathbf{Q}|$ which is characteristic for a zero range interaction with a finite value of the contact 
density $\mathcal{C}$ defined in~\eqref{eq:contact-Delta}. In the regime where an instability of the  FFLO type dominates, we accept a threshold of 
$\left|\hat{\Gamma}^{-1}\left(\mathbf Q, \Omega_n=0 \right)\right|\leq 10^{-6}$, for the dimensionless inverse vertex $\hat{\Gamma}^{-1}\equiv-(2\pi)^{3/2}\lambda_T^3\Gamma^{-1}/\beta$ at a finite value of $\mathbf Q$ as zero.
To check the precision of our numerical results, we compute the pressure from the grand potential in equation \eqref{eq:LW-functional} and compare it to the pressure obtained by the Tan 
pressure relation \eqref{eq:Tan-pressure}. The resulting relative deviations vary between a level of about $10^{-4}$ at generic points in the phase diagram to the $10^{-3}$ level 
close to the phase transition. Since this relation does not only depend on energy scales up to $k_F$, but is also very sensitive to the non-trivial high frequency and momentum behavior of both the vertex function and the Green functions, the extraordinarily high precision to which it is fulfilled exemplifies the ability of our specifically tailored methods for the Fourier transform to correctly capture slowly decaying functions over several orders of magnitude. Furthermore, in the case $h=0$ all our results coincide with the normal fluid computations in the previous Luttinger-Ward studies for the BCS-BEC crossover of spin-balanced Fermi gases \cite{haus07bcsbec}.\\

\begin{figure}[t]
\centering
\includegraphics[width=.95\columnwidth]{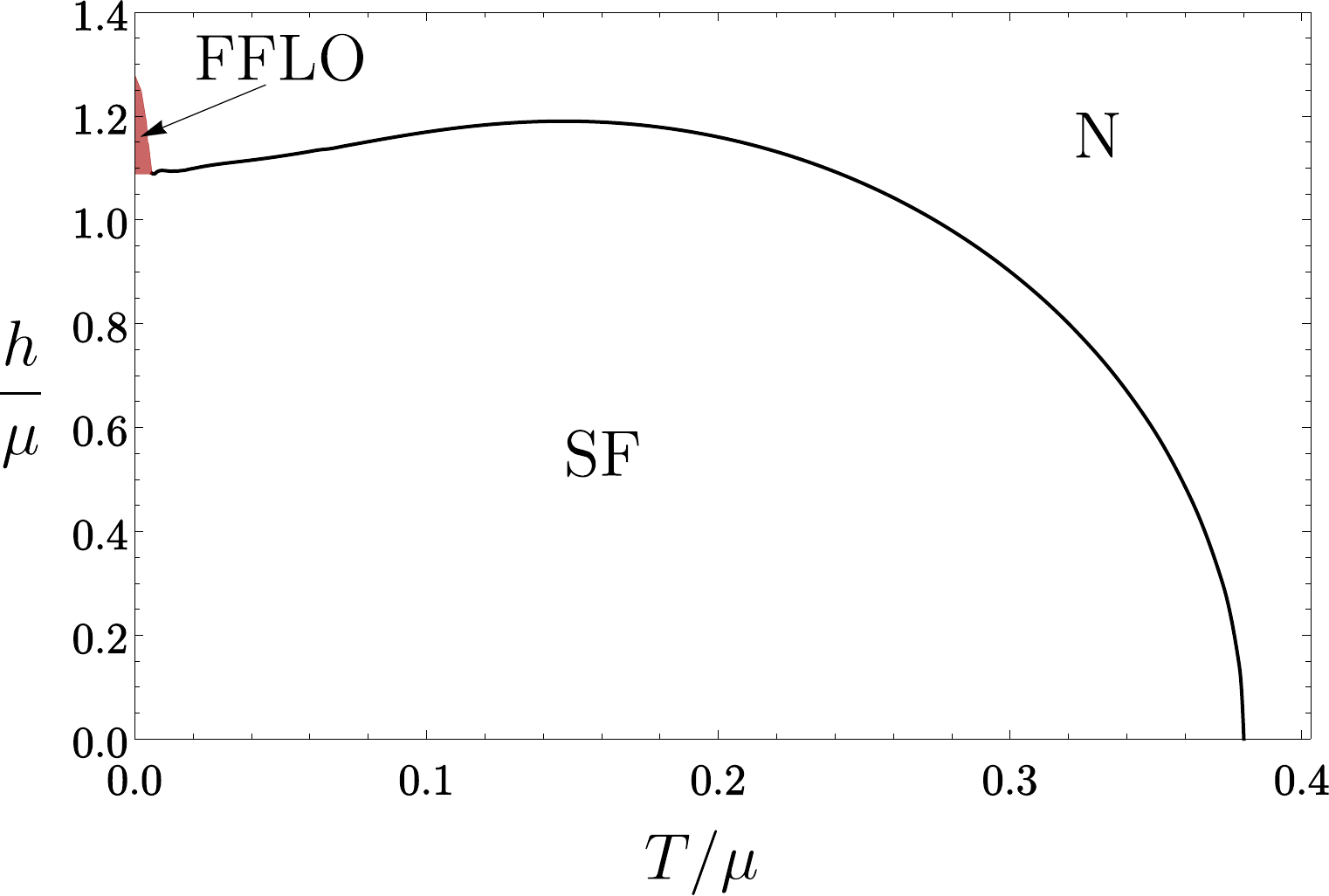}
\caption{(Color online) Black line: temperature dependence of the critical field strength $h_c$ at unitarity, below which the system forms a homogeneous superfluid. The extrapolation to zero temperature yields $(h/\mu)_c=1.09\pm0.05$. At very small temperatures and fields above $h_c$ FFLO order, here indicated by the red-shaded region, appears to emerge.} 
\label{fig:Tc_vs_h}
\end{figure}

The phase diagram of the unitary gas at finite temperature is shown in Fig.~\ref{fig:Tc_vs_h}.
The critical temperature at vanishing field $h=0$ is $(T/\mu)_c \simeq 0.38$, 
which corresponds to $T_c/T_F \simeq 0.15$. Apparently, at temperatures below
$T/\mu\simeq 0.15$  the critical field $(h/\mu)_c$ starts to decrease slightly.
The origin of this unexpected behavior is quite likely the presence of
a tricritical point, below which the normal to superfluid transition
is of first order. The present approach, which is restricted to the normal fluid, can only
determine the lower critical field, at which the normal state ceases to be at least metastable.
To faithfully determine the position of this tricritical point, and the critical field below it,
requires a calculation which includes the symmetry broken phase. This will be presented in future work.
At very low temperatures, we find a regime where the instability occurs at a finite wave vector.
An extrapolation to the lowest available temperatures shows that $(h/\mu)_{\rm FFLO} = 1.28 \pm 0.15$.
The precise value unfortunately depends on the criterion applied for the vanishing of the inverse
vertex at finite $Q$.\\

\noindent Our extrapolated value $(h/\mu)_c=1.09\pm 0.05$ for the CC-transition of the unitary gas is considerably larger than 
the one obtained by Lobo et al.~\cite{lobo06}, who find $(h/\mu)_c=0.96$, and also the result $(h/\mu)_c=0.83$ in the more recent work 
by Boettcher et al.~\cite{boet15}. Within leading order in a 
$1/N$--expansion, the result is 
$(h/\mu)_c=0.807\ldots$~\cite{niko07renorm}, while $(h/\mu)_c=1.947...$ in next-to-leading order \cite{veil07largeN}.  Our result
agrees quite well, however, with the one from the $\varepsilon=4-d$ expansion by Nishida and Son,
who obtain $(h/\Delta)_c=0.878$ to leading order~\cite{nish07eps}.  With $\Delta/\mu=1.31$
this gives $(h/\mu)_c=1.15$. Experimentally, the ratio has been inferred from in situ imaging 
of the density profiles in a trapped imbalanced gas by Shin et al.~\cite{shin08phasediagram}.  In particular,
it has been shown there that the parameter $\eta_c$ in $(h/\mu)_c=(1-\eta_c)/(1+\eta_c)$ can be
connected to the measured ratio $R_c/R_\shortuparrow$ of the radius $R_c$ of the balanced 
superfluid in the center of the trap and the outer radius $R_\shortuparrow$ of the fully polarized 
gas at the edge by the relation
\begin{equation}
\label{eq:MIT-experiment}
\eta_c=\frac{2\left[\xi_s\left(n_s(0)/n_0\right)^{2/3} -1\right]}{1-\left(R_c/R_\shortuparrow\right)^2} +1\, .
\end{equation}
Apart from the universal Bertsch parameter $\xi_s$, it only contains the ratio between the central density $n_s(0)$ 
of the balanced superfluid and the density $n_0$ of a fully polarized Fermi gas with chemical potential $\mu_\shortuparrow(0)$.
The measured values $R_c/R_\shortuparrow=0.43$ and  $n_s(0)/n_0=1.72$ for the coldest sample then give
$\eta_c=0.03$ or $(h/\mu)_c=0.95$ with a Bertsch parameter $\xi_s=0.42$. Using the precise result $\xi_s=0.37$, however, leads to a much larger value $(h/\mu)_c=1.35$ with the same set of experimental 
parameters. A possible explanation for the discrepancy with the result $(h/\mu)_c=1.1$ found here, is that the effective Bertsch parameter
at the finite temperature of the experiment is higher than the universal value at $T=0$. 
A rather small critical field $(h/\mu)_c=0.88$ has also been inferred in later measurements by Navon et al.~\cite{navo10thermo}, but again
the value extracted from the analysis of the density profiles changes substantially if the precise number of the Bertsch
parameter is used rather than the assumed value $\xi_s=0.42$.

\begin{figure}[t]
\centering
\includegraphics[width=\columnwidth]{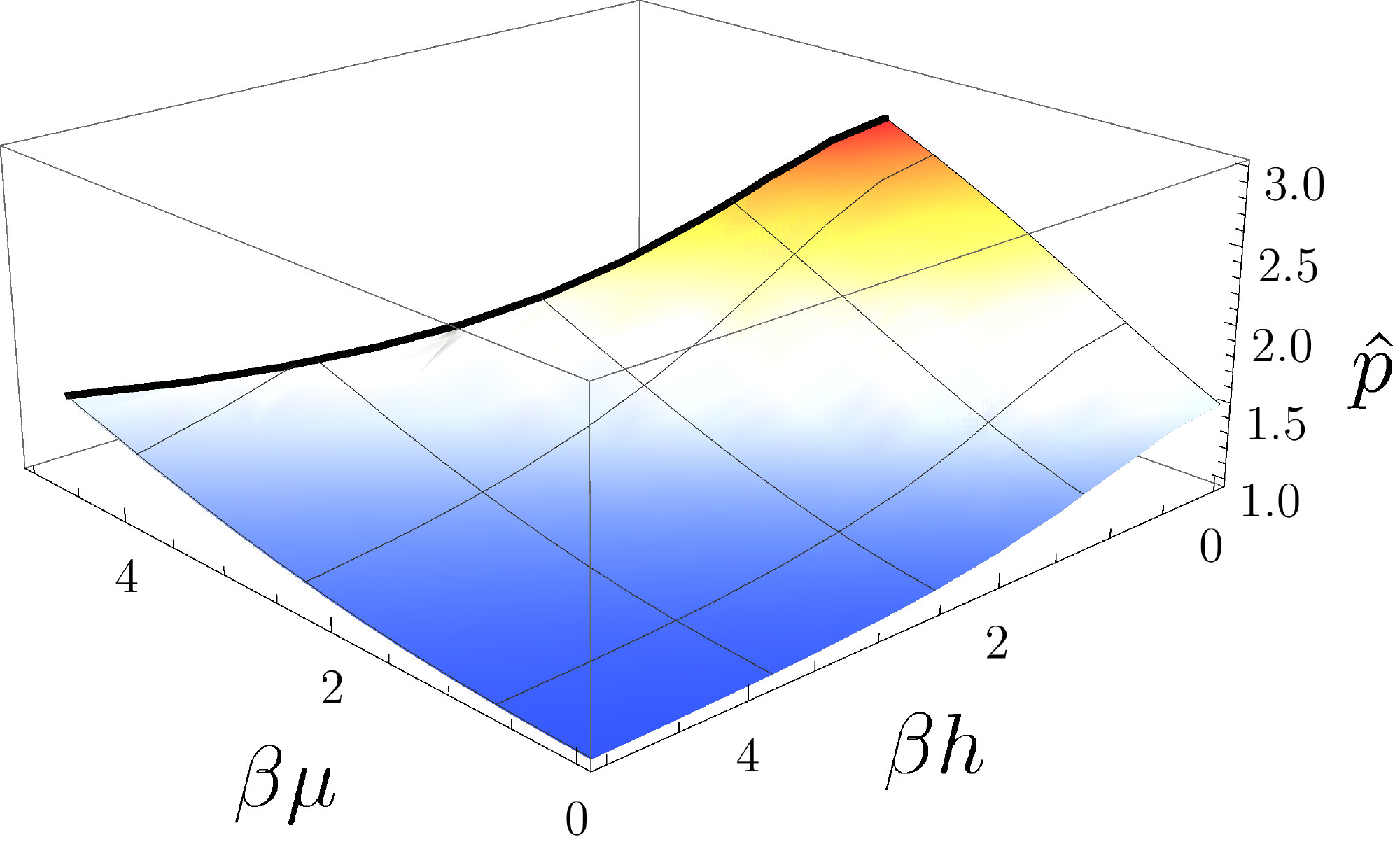}
\caption{(Color online) Dimensionless pressure scaling function $\hat{p}$ at unitarity as a function of the dimensionless chemical potentials. The bold black line indicates the superfluid transition.
} 
\label{fig:pressure}
\end{figure}

From the Luttinger-Ward functional \eqref{eq:LW-functional} one can directly determine the dimensionless pressure
 scaling function $\hat{p}=p(T,\mu,h,0)/(p^{(0)}(T,\mu_\shortuparrow)+p^{(0)}(T,\mu_\shortdownarrow))$ 
 which is shown in Fig.~\ref{fig:pressure}. At vanishing Zeeman field $h=0$ this reduces to $f_p(\beta\mu,0)$, 
 for which our approach yields $f_p((\beta\mu)_c\simeq2.65,0)\simeq 2.73$. Along the critical line, $\hat{p}$ has a 
 shallow minimum near $\beta h = 9$ (not shown) and slowly converges to $\simeq 1.5$ for $\beta h \to \infty$. 
 Far away from the critical line, for $h\gg\mu$, the system becomes strongly polarized and approaches the 
 noninteracting limit where $\hat{p}=1$.

\begin{figure}[t]
\centering
\includegraphics[width=\columnwidth]{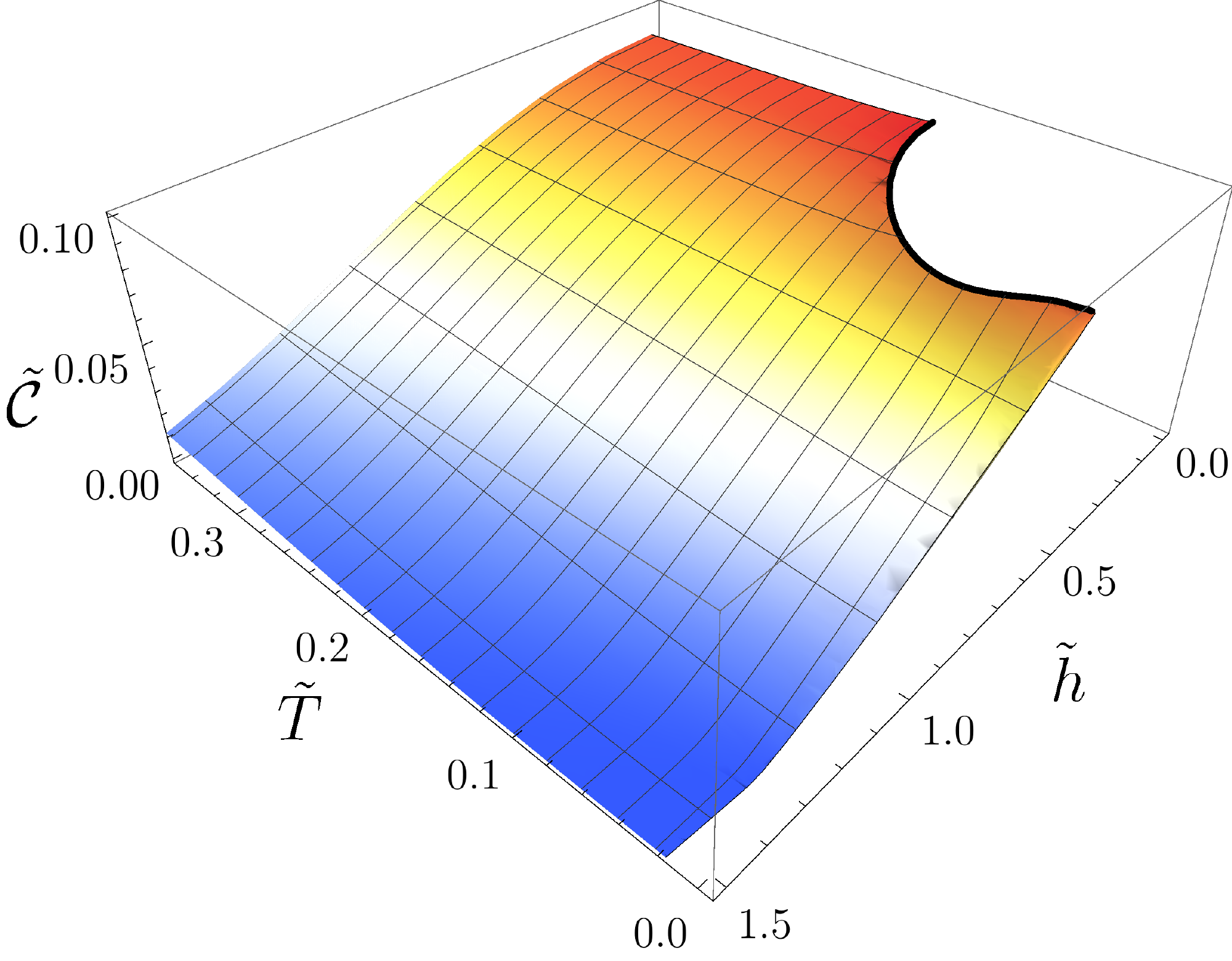}
\caption{(Color online) The dimensionless contact density $\tilde{\mathcal{C}}=\mathcal{C}/k_F^4$ at $1/a=0$ as a function of the reduced variables $\tilde{T}=T/T_F$ and $\tilde{h}=h/\varepsilon_F$ decreases strongly with increasing Zeeman field. The system becomes unstable to pair formation at the black line.} 
\label{fig:Contact}
\end{figure}

The intensive Tan contact density $\mathcal{C}$, which appears in the tail of the momentum distribution 
$n_{\sigma}(\mathbf{k})\to\mathcal{C}/k^4$~\cite{tan08a, haus94} at large momenta can be obtained 
from the vertex function via \eqref{eq:contact-Delta} or also from the pressure, using 
\begin{align}
\frac{\hbar^2}{4\pi m}\mathcal{C}=\frac{\partial p}{\partial\left(1/a\right)}\, .
\end{align}
The equation of state has to obey the exact Tan relation~\cite{tan08b}
  \begin{equation}
p=\frac{2}{3}\,\varepsilon+\frac{\hbar^2}{12\pi ma}\,\mathcal{C}\,  ,
\label{eq:Tan-pressure}
\end{equation}
which connects pressure $p$ and energy density $\varepsilon$. 
In particular, at unitarity, $p=2\,\varepsilon/3$ as a consequence of scale invariance.  
Quantitative results for the contact density at unitarity
are presented in Fig.~\ref{fig:Contact}. It has a maximum value $\tilde{\mathcal{C}}\simeq 0.09$ 
near the  superfluid transition at $\tilde{h}_c$, which is essentially independent of the specific value $\tilde{h}_c$ of the field. 
For $\tilde{h}$ beyond $\tilde{h}_c$, the imbalance grows and the contact density $\tilde{\mathcal{C}}$ decreases quickly. 
Indeed, in the limit $n_{\shortdownarrow}\to 0$ of a strongly imbalanced gas, the contact density
has to vanish because there is no tail in the momentum distribution for a single-component,
non-interacting Fermi gas.  More precisely, $\mathcal{C}\sim \tilde{s}\, {k_F}_\shortuparrow n_{\shortdownarrow}$ 
turns out to vanish linearly with the minority density $n_\shortdownarrow$, where ${k_F}_\shortuparrow$ is the 
Fermi wave vector of the majority component and  $ \tilde{s}$ a dimensionless prefactor of order one~\cite{punk09molaron}. 
This behavior is in fact captured correctly by our results. Note also that, at a given value of the field $h$,
the contact density is largely independent of the temperature in the relevant regime $\tilde{T}\lesssim 0.3$.

\begin{figure}[t]
\centering
\includegraphics[width=\columnwidth]{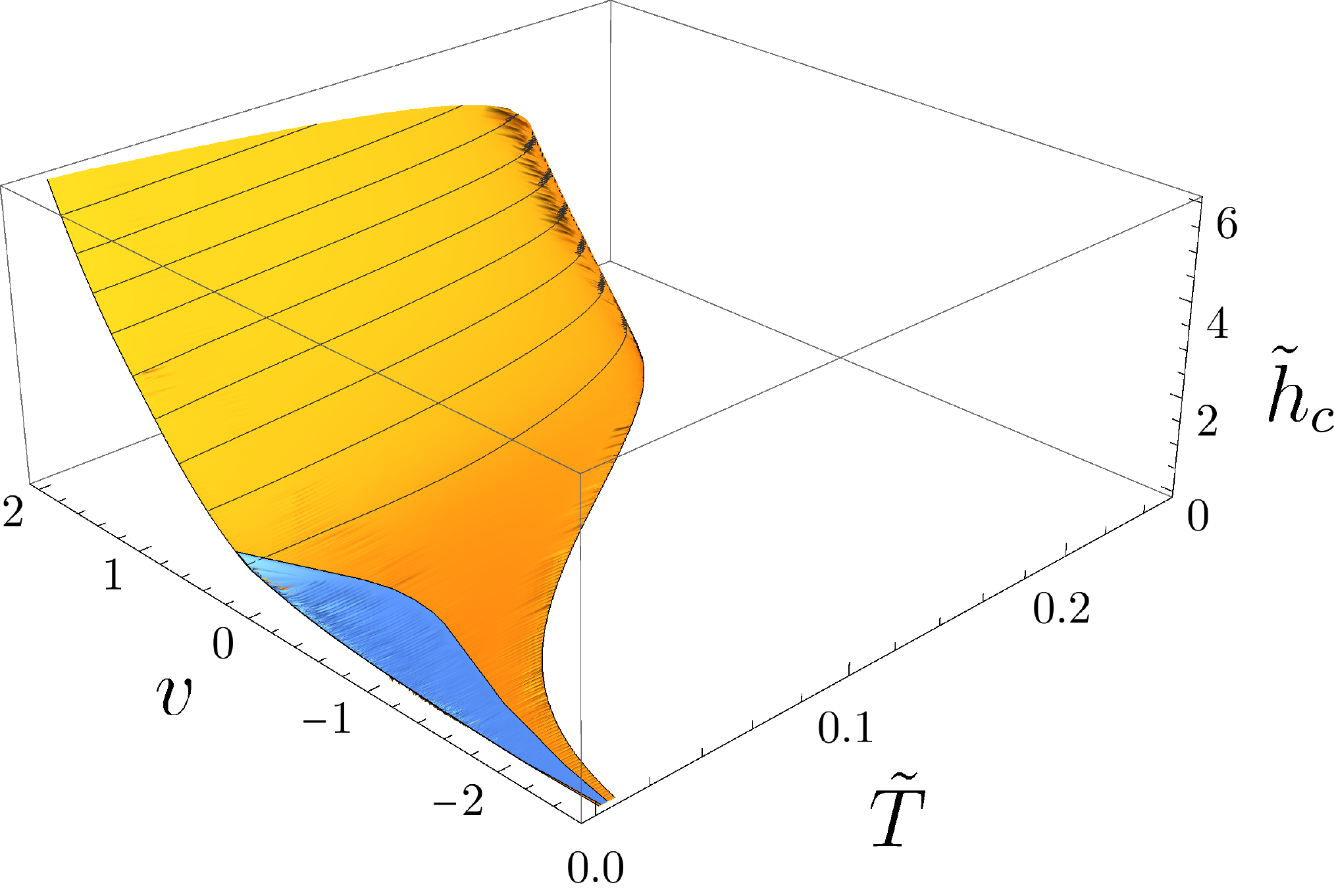}
\caption{(Color online) Critical Zeeman field strength $(h/\varepsilon_F)_c$ as a function of temperature $T/T_F$ and inverse scattering length $v$. Below the yellow surface the system forms a homogeneous superfluid. Underneath the blue wedge an FFLO phase is predicted. Near unitarity critical temperatures for FFLO order reach as high as $T_c/T_F\simeq0.03$.} 
\label{fig:Tc3D}
\end{figure}

Finally, in Fig.~\ref{fig:Tc3D}, we show the critical field strength $\tilde{h}_c=(h/\varepsilon_F)_c$ as a function of the reduced temperature $\tilde{T}=T/T_F$ 
and the dimensionless interaction strength $v=1/(k_F a)$ in the crossover regime $|v|<2$. While the line of critical temperatures for the balanced gas 
$T_c(\tilde{h}=0)/T_F$ coincides with previous results, we generally observe decreasing values of the critical temperatures with growing Zeeman field. 
The FFLO wedge extends from the BCS limit to the unitary regime with a maximum in the transition temperature on the order of $0.03\,T_F$ for 
interaction strengths $v \simeq -0.75$ slightly on the BCS side of the unitary limit. 

\section{CONCLUSION}  
\label{sec:conclusion}

We have reanalyzed the basic problem of Fermi gases with a finite Zeeman field.
In the context of ultracold atoms, where the relevant range is the one near 
infinite scattering length, quantitatively reliable results for the thermodynamic functions
at finite temperature for this problem are still rare. Based on an extension of an 
earlier Luttinger-Ward approach to finite imbalance, our results provide strong evidence   
for the presence of an FFLO phase for Fermi gases near unitarity, consistent with 
the rather general, non-perturbative argument given by Son and Stephanov~\cite{son06imbalance}. 
The associated critical temperatures are unfortunately rather small, with a maximum 
of order $\simeq 0.03\, T_F$ slightly on the BCS side of the unitary gas. In addition, we have provided explicit 
results for the universal scaling functions and the contact density of the imbalanced 
gas in its normal state. This analysis is based on understanding the thermodynamics 
of imbalanced Fermi gases near unitarity in terms of universal scaling functions near 
the strong coupling fixed point at zero density discussed first by Nikoli\'{c} and Sachdev~\cite{niko07renorm}.
Of course, a detailed analysis of the various symmetry broken phases is still open.
In particular, it seems important to study both the precise position and the
thermodynamic phases near the splitting point, where the quasiparticle dispersion 
of the balanced superfluid changes its nature.  

{\bf Acknowledgement} This paper is dedicated to Lev Pitaevskii, to whom W.Z. is grateful for
discussions and insightful remarks on the physics of ultracold atoms and beyond during many years.
We also acknowledge financial support of this work by the Nano-Initiative Munich (NIM).

\bibliographystyle{plain}
\bibliography{References}




\end{document}